\documentclass[aps,prb,twocolumn]{revtex4}
\usepackage{boxedminipage}
\usepackage{lscape}
\usepackage{float}
\usepackage[ansinew]{inputenc}
\usepackage{epsfig}
\usepackage{subfigure}
\usepackage{graphicx}
\usepackage{amsmath}
\usepackage{amssymb}
\usepackage{amsthm,url}
\usepackage[french]{babel}  

\renewcommand{\d}[2]{\frac{#1}{#2}}
\newcommand{\pd}{\partial}

\newcommand{\barray}{\begin{eqnarray}}
\newcommand{\earray}{\end{eqnarray}}

\newcommand{\beq}{\begin{equation}}

\newcommand{\eeq}{\end{equation}}

\begin{document}

\title{Entropy, frustration and large thermopower of doped Mott insulators on the fcc lattice}
\author{Louis-Fran\c{c}ois Arsenault$^{1}$, B. Sriram Shastry$^{2}$, Patrick S\'{e}mon$^{1}$, A.-M. S. Tremblay$^{1,3}$}
\affiliation{$^1$ D\'{e}partement de Physique and RQMP, Universit\'{e} de Sherbrooke, Sherbrooke,
QC, Canada\\
$^{2}$ Physics Department, University of California, Santa Cruz, CA 95064,
USA\\
$^{3}$Canadian Institute for Advanced Research, Toronto, Ontario, Canada.}
\date{\today}

\begin{abstract}
Electronic frustration and strong correlations may lead to large Seebeck coefficients. To understand this physics on general grounds, we compute the thermopower of the one-band Hubbard model on the 3-dimensional fcc lattice over the whole range of fillings for intermediate and large interaction strength. Dynamical mean-field theory shows that when the density approaches half-filling, the fcc lattice at strong coupling exhibits a large low temperature Seebeck coefficient $S$. The largest effect occurs as one approaches $n=1$ from dopings where electronic frustration is maximized. The high-frequency limit of the thermopower and the Kelvin limit are both used to provide physical insight as well as practical tools to estimate the thermopower. The high-frequency limit gives a reliable estimate of the DC limit at low temperature when the metal becomes coherent. By contrast, the Kelvin approach is useful in the strongly interacting case at high temperature when transport is incoherent. The latter result shows that in doped Mott insulators at high temperature and strong coupling the thermopower can be understood on entropic grounds.
\end{abstract}

\maketitle

\hyphenation{Brillouin}


\section{Introduction}
The search for better thermoelectric materials has focussed on strongly correlated materials in recent years, going beyond traditional semiconductor systems. Narrow-band correlated oxides in quasi two dimensional systems such as sodium cobaltate $Na_xCoO_2$ \cite{terasaki}  show large thermoelectric power (or Seebeck coefficient $S$) $S\sim 100 \mu V/K$, at $\sim 100K$. Mott Hubbard physics is invoked to understand this extraordinary scale of $S$ in $Na_{x}CoO_{2}$ and $Sr_{x}La_{1-x}TiO_{3}$ \cite{Mukerjee:2007}, often in the limit of zero bandwidth or atomic limit $t\rightarrow 0,$ \cite{beni,Mukerjee:2007,Mukerjee:2005}. More recently, $Fe Sb_2$ (a Kondo insulator or a correlated band insulator) has drawn much attention.~\cite{bentien,Sun,Tomczak:2010} This is a three-dimensional compound with a colossal Seebeck coefficient $S\sim~45,000 \ \mu V/K$ that is of great interest for thermoelectric cooling at cryogenic temperatures.

For more traditional classes of thermoelectric materials, such as the  semiconductor $Bi_2Te_3$, the Bloch-Boltzmann transport theory points to a way to optimize the  band structure for thermoelectric applications \cite{Mahan:1996}. In the case of strongly correlated materials,  the situation is much more complicated due to the fragility of the quasi-particles. These appear with strongly renormalized physical properties only in the so-called coherent regime that may develop at very low temperatures in the presence of correlations. Strongly correlated systems favour large thermopower then because the large effective masses correspond to small Fermi temperatures. The temperature scale for maximum thermopower is set by fractions of hopping $t$. Estimates for real materials give for $V_{2}O_{3}$ \cite{Rozenberg:kotliar:Kajueter} $t\approx 580K$, for $FeSi$, \cite{Sluchanko:Glushkov:Demishev:Kondrin:Petukhov:Samarin:Moshchalkov:Menovsky},  $t\approx 152K$, and $t\approx 100K$ for Na$_{x}$CoO$_{2}$ ~\cite{Haerter:Peterson:Shastry}. Hence the  peaks that are found for $T<t$ as well as the strong temperature dependence up to about $T\sim t$  are in a physically relevant temperature range not far from room temperature. Room temperature may even fall in the incoherent regime where Fermi liquid theory does not apply and $S$ can be large.

Early studies of the effect of correlations on the thermopower focused on the Hubbard model on unfrustrated lattices~\cite{Schweitzer:1991,Pruschke:1995,Palsson:Kotliar:1998,Oudovenko:2002,Grenzebach:2006}. On these unfrustrated lattices, it is unlikely that large values of thermopower $S$ can be achieved~\cite{Haule:Kotliar:2009} because when the temperature $T$ is of the order of the coherence temperature $T^{coh}$, the thermopower has contributions from the quasiparticle peak but also from the adjacent Hubbard band and these contributions have opposite sign. Study of the effect of frustration on the cubic lattice with next-nearest-neighbor hopping at half-filling for interaction strengths below the Mott transition~\cite{Merino:2000} show that the peaks that are obtained in $S$ are similar to those in the electronic specific heat and are associated with the thermal destruction of the quasiparticles. More recently, realistic calculations~\cite{Oudovenko:2006,Haule:Kotliar:2009} of thermopower for experimentally interesting compounds have shown increase in thermopower due to correlations~\cite{Arita:2008,Held:2009,Tomczak:2010}. The role of frustration has been emphasized in particular for the cobaltate $Na_xCoO_2$ using the $t-J$ model on a 2d finite triangular lattice~\cite{Haerter:Peterson:Shastry}. Realistic calculations for that compound~\cite{Wissgott:Toschi:Sangiovanni:Held:2011,Wissgott:Toschi:Usui:Kuroki:Held:2010} demonstrate that frustration and interactions combined also with disorder can lead to large enhancements.

Here we focus on a generic model system with no disorder. More specifically, we study the fcc lattice because of its prevalence in materials and because its electronic structure is frustrated already for nearest-neighbor hopping. Shastry and coworkers \cite{shastry_review,peterson_1,peterson_2,peterson_shastry} have argued that in correlated materials, electronic frustration can amplify the Seebeck coefficient considerably beyond Bloch-Boltzmann (band) theory expectations. The fcc lattice has optimal thermopower, according to the Bloch-Boltzmann theory \cite{Mahan:1996}, when the chemical potential coincides with the sharp peak in the density of states. Here we show, instead, that in the presence of interactions, large thermoelectric effects  appear at low temperatures close to half-filling and for dopings that correspond to large electronic frustration, provided the system is sufficiently correlated that the half-filled band is insulating.  We show that insights on the temperature and doping dependence of our results may be obtained from methods developed recently by Shastry and coworkers \cite{shastry_review,peterson_1,peterson_2,peterson_shastry}. In these approaches, one considers either the high-frequency limit or the ultra-low frequency (Kelvin) limit. In the latter case, the thermopower is related to the density dependence of the entropy at fixed temperature. We find that such entropic considerations beyond the Bloch-Boltzmann theory can control the thermopower in the incoherent regime and that in the coherent regime the high-frequency limit is often reliable.  In other words, while approximate, these approaches also give correct order-of-magnitude estimates and bypass the difficulties facing Kubo formula calculations. These methods can be straightforwardly implemented in existing codes for realistic calculations.~\cite{Xu_Weber_Kotliar:2011}

In the next section we describe the model and the Dynamical mean-field theory (DMFT) approach along with the three formulas used for thermopower: Kubo, high-frequency and Kelvin. The results are then discussed for two values of interaction strengths, half bandwidth and two times the bandwidth, in the context of the Kubo formula. The next to last section makes comparisons with the high-frequency and Kelvin approaches. We end with a summary. Appendix A shows that in the strong-coupling limit the thermopower changes sign at half-filling $n=1$ instead of $n=0.8$ in the Bloch-Boltzmann theory. Appendix B presents benchmarks, in the worse case, of the impurity solver we used for the Kubo calculations.

\section{Model and methods}
\subsection{DMFT}
Dynamical mean-field theory (DMFT) has proven its accuracy for the Hubbard model and its predictive power for real materials \cite{dmft,Kotliar:2006} that have a clearly three-dimensional structure, despite the fact that the method is exact only in infinite dimension. Here we apply it to the one-band Hubbard model on  the fcc lattice where the single-particle dispersion is given by $\varepsilon _{\mathbf{k}}=-4t[\cos (k_{x})\cos(k_{y})+\cos (k_{x})\cos (k_{z})+\cos (k_{y})\cos (k_{z})]$. We take nearest-neighbor hopping $t$ positive.

We consider only the paramagnetic phase. To solve the impurity problem of DMFT, we use two different methods. 1) The numerically exact continuous time quantum Monte Carlo method (CTQMC) \cite{Werner:2006}, a finite temperature approach that relies on the Monte Carlo summation of all diagrams obtained from the expansion of the partition function  in powers of the hybridization $\Delta $. This method does not have errors associated with time discretization or bath parametrization and is therefore exact within statistical errors, but computationally expensive. When analytical continuation is an issue for this method, we rely on: 2) Iterated Perturbation Theory (IPT), an approximation method based on an interpolation from $2^{nd}$ order perturbation theory for the Anderson impurity problem \cite{Kajueter:kotliar}. The interpolation preserves the correct high-frequency limit for the self-energy and is exact in both the non-interacting and the atomic limits. For $U/t$ in the intermediate-coupling regime, the condition of Refs.~[\onlinecite{Martin:1986}] and [\onlinecite{Merino:2000}] to fix the bath occupation number leads to accurate results. However, at finite temperature and large $U$, we need to use the version of IPT that was developed and carefully benchmarked in Ref.~[\onlinecite{Arsenault:Semon:Tremblay:2012}]. In this approach, IPT-$D$, one needs double occupancy. At very large $U$, the simplest estimate for double occupancy suffices. We checked for many observables that CTQMC and IPT in this particular implementation give similar temperature dependent properties.

\subsection{Thermopower}  A general Kubo formula for the thermopower may be written down using linear response theory \cite{shastry_review} as
\begin{equation}
S(q_{x},\omega )=\frac{\chi _{{\hat{J}}_{x}(q_{x}),K(-q_{x})}(\omega )}{T\
\chi _{{\hat{J}}_{x}(q_{x}),\rho (-q_{x})}(\omega )},  \label{General_Kubo}
\end{equation}
where the susceptibility of any two operators $A,B$ is given by $\chi_{A,B}(\omega )=i\int_{0}^{\infty }\ dt\ e^{i\omega t-0^{+}t}\langle \left[
A(t),B(0)\right] \rangle $ and where $\rho $, $K=H-\mu \hat{N}$ and $\hat{J}_{x}$ are the charge density, the (grand canonical) Hamiltonian and the
current operator respectively at finite wave vectors. The \emph{fast} limit where $\mathbf{q}\rightarrow 0$ first and
then $\omega \rightarrow 0$,$\;$ is the relevant limit, that we will call $S_{Kubo}$, for DC transport.~\cite{luttinger} In the DMFT\ limit the vertex corrections vanish for one band models \cite{Paul:2003,Freericks:2001} but one still needs analytical continuation of single-particle Green's functions since the methods of solution for DMFT are formulated in Matsubara frequency.  That may be done to obtain meaningful results in the simplest cases~\cite{Merino:2000}. However, the most accurate modern method to solve the DMFT equations, namely CTQMC, requires analytical continuation of numerical data that contains statistical uncertainties. The fact that the kernel for $\chi _{{\hat{J}}_{x}(q_{x}),K(-q_{x})}(\omega )$ is not positive definite renders this in practice extremely difficult, even for maximum entropy methods that are usually called for with CTQMC results. That is why to compute $S_{Kubo}$ we performed the analytical continuation on the self-energy obtained by IPT-$D$ using the Pad\'{e} N points algorithm that can be found in Ref.~[\onlinecite{Vidberg:1977}]. $S_{Kubo}$ is then calculated using the real frequency equations~\cite{Merino:2000}. In Ref~[\onlinecite{Arsenault:Semon:Tremblay:2012}], we showed that the results obtained for the resistivity were at least qualitatively valid.

Two new approximate methods to compute the thermopower  \cite{shastry_review,peterson_shastry} without analytical continuation, namely the \emph{superfast limit} $S^*$ and the Kelvin formula $S_{Kelvin}$, are explained below. They have the advantage that they do not require analytical continuation. In addition, they have physical content that will help us identify the origin of enhanced thermopower.

Consider first the superfast limit of Eq.(\ref{General_Kubo}), $S^{\ast }=\lim_{\omega \gg \omega _{c}\;\;\vec{q}\rightarrow 0}\ S(q,\omega
),$ where $\omega _{c}$ is the largest characteristic frequency in the problem. In the case of the Hubbard model, we would identify $\omega _{c}=\mbox{max}\{W,U\}$, where $W$ is the bandwidth. In terms of the thermoelectric operator $\Phi ^{xx}$ and the stress  tensor $\tau ^{xx}$ that appears in the f-sum rule~\cite{shastry_review}, the quantity $S^{\ast }$ is equal $\frac{\langle \Phi ^{xx}\rangle }{T\langle \tau ^{xx}\rangle }$.
Since the Seebeck coefficient is a ratio of transport coefficients where the scattering rate cancels in the relaxation time approximation, it is conceivable that it may have a weak dependence on frequency. It turns out that $S^*$ captures much of the many-body content of the exact $S_{Kubo}$ when the characteristic $\omega _{c}$ is not too large.~\cite{shastry_review} For the Hubbard model, the commutators can be evaluated exactly and the correlation functions evaluated solely in terms of the single-particle Green's function~\cite{shastry_unpub,Xu_Weber_Kotliar:2011}
\begin{eqnarray}
\langle \Phi ^{xx}\rangle =\frac{q_{e}}{\beta }\sum_{\mathbf{k},n,\sigma
}e^{i\omega _{n}0^{+}}&&G_{\sigma }(\mathbf{k},i\omega _{n}) \Bigg\{\Sigma
_{\sigma }(\mathbf{k},i\omega _{n})\ \frac{\partial ^{2}\varepsilon _{%
\mathbf{k}}}{\partial k_{x}^{2}}  \notag  \label{final_phistar} \\
&&+\frac{\partial }{\partial k_{x}}\ \left( \frac{\partial \varepsilon _{%
\mathbf{k}}}{\partial k_{x}}(\varepsilon _{\mathbf{k}}-\mu )\right) \Bigg \},
\notag
\end{eqnarray}%
with $q_{e}$ the (negative) charge of the electron. While one can show that $\lim_{T\rightarrow 0}\langle \Phi ^{xx}\rangle \ =0$ when $U=0,$ for interacting systems it was found in numerical calculations \cite{peterson_1,peterson_2,Xu_Weber_Kotliar:2011} and from Fermi liquid theory that $\lim_{T\rightarrow 0}\langle \Phi ^{xx}\rangle \ \neq0$ and hence, $S^{\ast}$ diverges as $T$ vanishes. Indeed, at large frequency there are reactive parts to the energy transport that should not be considered part of the thermal current. We thus eliminate the constant term $\lim_{T\rightarrow0}\langle \Phi ^{xx}\rangle \equiv \langle \Phi ^{xx}\rangle _{0}\ $by fitting the low temperature part of $\langle \Phi ^{xx}\rangle $ to $\langle\Phi ^{xx}\rangle _{0}+bT^{2}$, a functional form derived from the Sommerfeld expansion that empirically remains valid in the presence of strong interactions. We then define the thermopower in the super fast limit by
\begin{equation}
S^{\ast \ast }=\frac{1}{{T}}\frac{{\langle \Phi ^{xx}\rangle -}\langle \Phi
^{xx}\rangle _{0}}{{\langle \tau ^{xx}\rangle }}.  \label{S**}
\end{equation}
Not subtracting $\langle \Phi^{xx}\rangle _{0}$ leads to large deviations between $S^{\ast \ast }$ (now only $S^{\ast}$) and $S_{Kubo}$ at low $T$ for all dopings, contrary to what we find.~\cite{Xu_Weber_Kotliar:2011}

The second approximate method considers the slow limit of the general formula for thermopower, Eq.~\eqref{General_Kubo}. As noted earlier \cite{shastry_review,peterson_shastry}, this leads to a finite and interesting answer that is termed the Kelvin formula
\begin{equation}
S_{Kelvin}=\lim\limits_{\substack{\vec{q}\rightarrow 0 \\ \omega \rightarrow 0}}S(q,\omega)= \frac{k_{B}}{q_{e}}\left( \frac{\partial s}{\partial n}\right) _{T,V}=-\frac{k_{B}}{q_{e}}\left( \frac{\partial \mu }{\partial T}\right) _{V,n}
\label{skelvin}
\end{equation}
where $s$ is the entropy density, $n$ the number density, $V$ the volume, $\mu$ the chemical potential, $k_B$ Boltzmann's constant and where the last form follows from a Maxwell relation. The formulas for $S^{\ast \ast }$ and $S_{Kelvin}$ approach $S_{Kubo}$ from two different limits. \newline

\begin{figure}[tbp]
\begin{center}
\includegraphics[scale=0.6,bb= 101   285   509   540]{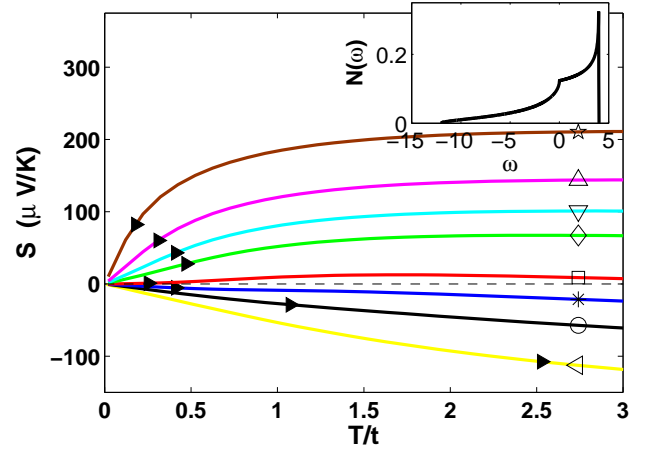}
\end{center}
\caption{(Color online) $S$ in $\protect\mu V/K$ for $U=0$ as a function of $%
T/t$ for different values of density: $n=0.2$ (yellow ($\vartriangleleft $)) $0.4$ (black ($\circ $)), $0.6$ (blue ($\ast $)), $0.8$ (red ($\square $)), $1.2$ (green ($\lozenge $)), $1.4$ (cyan ($\triangledown $)), $1.6$ (magenta ($\vartriangle $)) and $1.8$ (brown ($\star $)).}
\label{graph_U0}
\end{figure}

\begin{figure}
\begin{center}
\mbox{\includegraphics[scale=0.5,bb= 96   238   515   550]{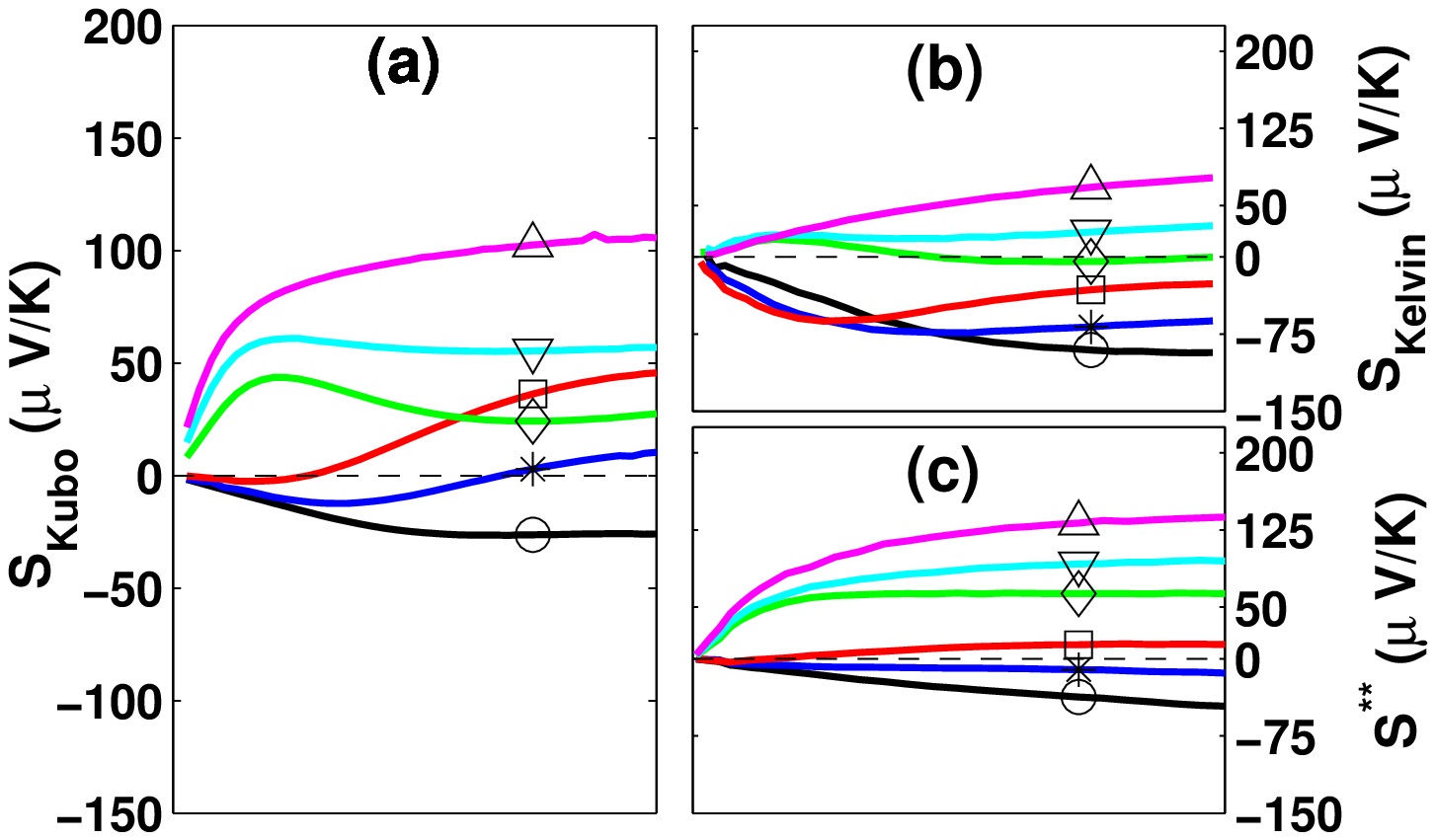}}
\mbox{\includegraphics[scale=0.5,bb= 96   280   515   521.5]{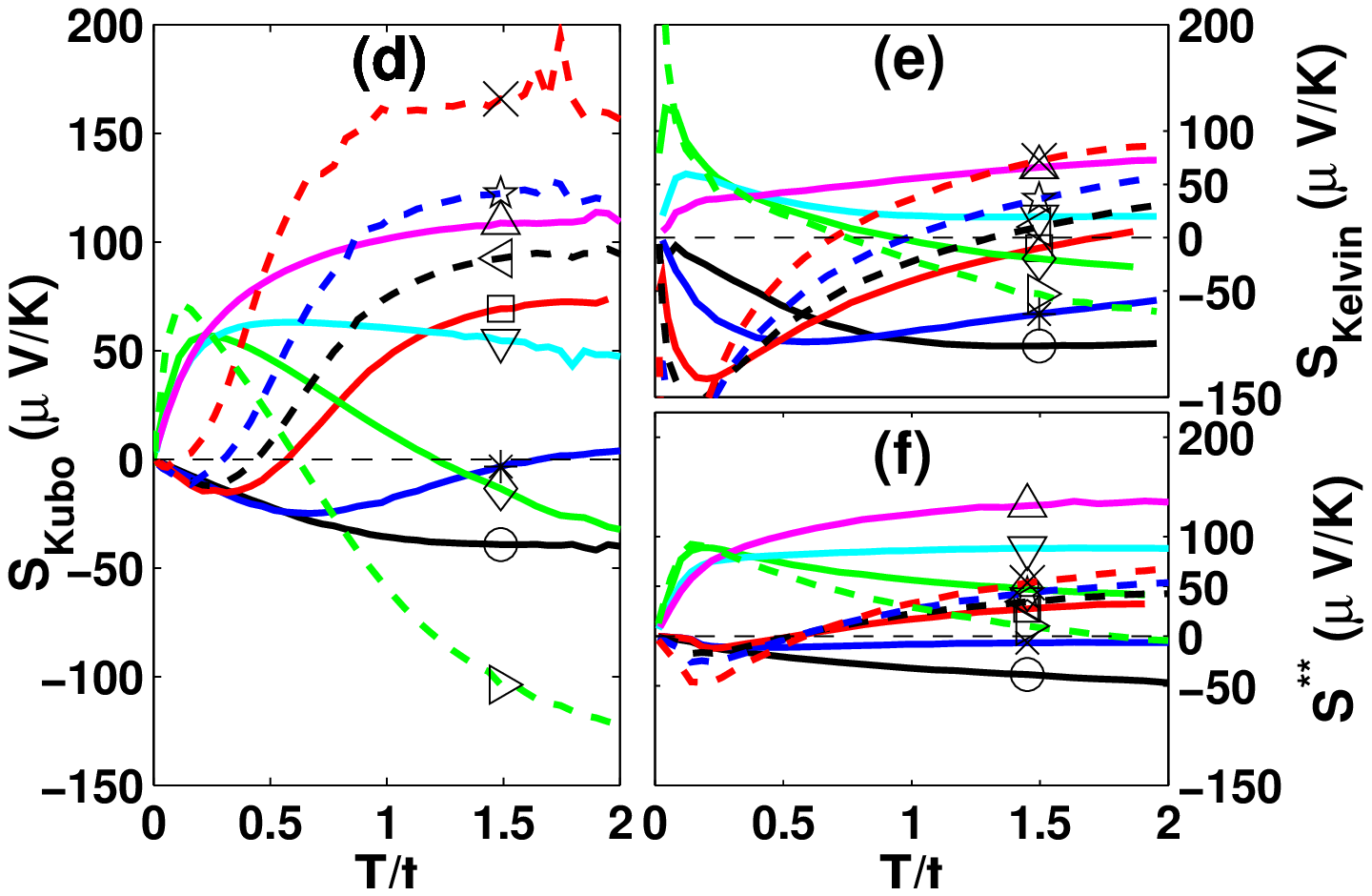}}
\end{center}
\caption{(Color online) $S_{Kubo}$, $S_{Kelvin}$ and $S^{**}$ as a function of $T/t$ as calculated with
IPT for Kubo and CTQMC for the Kelvin and high-frequency estimates for two values of interaction $U=8t$ (a), (b), (c) and $U=32t$ (d), (e), (f) for
different values of density: $n=0.4$ (black ($\circ $)), $0.6$ (blue ($\ast$), $0.8$ (red ($\square $)), $0.84$ (black - - ($\vartriangleleft $)), $0.88$ (blue - - $\star$), $0.92$ (red - - ($\times $))), $1.08$ (green - - ($\triangleright$)) $1.2$ (green ($\lozenge $)), $1.4$
(cyan ($\triangledown $)) and $1.6$ (magenta ($\vartriangle $)).}
\label{fig:S_Kubo_IPT}
\end{figure}

\section{Results}
\subsection{Band limit}
The Seebeck coefficient in the band limit $\left( U=0\right) $ and constant relaxation-time approximation is displayed in Fig.~\ref{graph_U0} along with the non-interacting density of states as an inset. The bare bandwidth is $16t$. The horizontal triangles $\vartriangleright$ on this plot indicate the temperature $T_{\ell }$, below which the leading  term of the Sommerfeld expansion for the $U=0$ case is $90\%$  of the full answer. The sign and magnitude of the results in Fig.~\ref{graph_U0} depend mostly on the particle-hole asymmetry $(X (\omega)\neq X (-\omega))$ of the energy-dependent transport function
\begin{equation}
X (\omega) = \sum_\mathbf{k} \left(\d{\pd\varepsilon_\mathbf{k}}{\pd k_x}\right)^2\delta (\omega-\varepsilon_\mathbf{k})
\label{TransportFunction}
\end{equation}
computed in Ref.~[\onlinecite{Arsenault:Semon:Tremblay:2012}]. At the qualitative level, the behavior of the thermopower is easy to understand. Below  $T_{\ell}$ it is linear, the behavior expected from Fermi liquid theory. The low temperature thermopower changes sign from negative to positive when the Fermi surface turns from electron-like to hole-like at the van Hove singularity in the density of states, located around $n=0.8$. In the large temperature limit, the thermopower saturates to a value of the order of the fundamental unit of thermopower, $k_B/e=86 \mu V/K$. As can be seen from the sharp peak in the density of states, the absolute value of $S$ is maximum in the large $T$ limit and for an almost filled band where the largest deviations from particle-hole symmetry occur. This asymmetry comes from the fact that the electronic spectrum for large filling is frustrated: the extrema of the band dispersion for excitations are not at high-symmetry points in the Brillouin zone. This may be understood simply from the electronic spectrum of a single triangle. Any site forms, with any pair of its nearest-neighbors on the FCC lattice, an equilateral triangle. Note that frustration leads to low values of $T_{\ell }$.

\subsection{Weak and strong coupling limits}
Armed with this understanding, we turn to the two interacting cases displayed in Fig.~\ref{fig:S_Kubo_IPT}. The value $U=8t$, namely half the bandwidth in Figs.~\ref{fig:S_Kubo_IPT}(a),(b),(c), is in the intermediate coupling range whereas $U=32t$, twice the bandwidth, in Figs.~\ref{fig:S_Kubo_IPT} (d),(e),(f) is representative of strong coupling. These values of $U$ are respectively below and above the Mott transition at half-filling. For each value of $U$ we evaluate $S$ in three different ways: $S_{Kubo}$, $S_{Kelvin}$ and $S^{**}$.

Let us move slowly through the many different results displayed in Figs.~\ref{fig:S_Kubo_IPT}. In these figures we do not display the results far away from half-filling, namely $n=0.2$ and $n=1.8$, since they are essentially the same as in the band limit Fig.~\ref{graph_U0}. Indeed, at low density the carriers (electrons at $n=0.2$ and holes at $n=1.8$) can avoid each other and the interaction is not important. Raising the carrier densities, we see that for $n=0.4$ (black) and $n=1.6$ (magenta) the results still do not depend much on $U$. In addition, the results for $S_{Kubo}$ and $S^{**}$ are very close to each other.

As we move closer to half-filling, the thermopower becomes more sensitive to interactions. Let us then first focus on $S_{Kubo}$  Fig.~\ref{fig:S_Kubo_IPT}(a) at weak to intermediate coupling, $U=8t$. At densities $n=0.6$ (blue) $n=0.8$ (red) and $n=1.2$, (green) $n=1.4$ (cyan) the main qualitative changes are a sign change as a function of $T$ for $n=0.6$ (blue) $n=0.8$ (red) and the appearance of a small low temperature peak for the four fillings. However, the absolute values do not exceed the largest absolute values that can be reached in the band limit by doping.

At strong coupling, $U=32t$, there are larger deviations from the $U=0$ results. The system is an insulator at $n=1$. Upon doping, it develops a large Seebeck coefficient with non-trivial $T$ dependence even below $T_{\ell }$ defined in Fig.\ref{graph_U0}. In Fig.~\ref{fig:S_Kubo_IPT}(d), at the fillings discussed above, $n=0.6$ (blue) $n=0.8$ (red) and $n=1.2$, (green) $n=1.4$ (cyan), the peaks in $S_{Kubo}$ are more pronounced at large $U$ and there is a sign change at finite temperature even for $n=1.2$. The maximum magnitude of thermopower is obtained at low temperature: for example for $n=0.8$ (red), $S \approx -15 \mu V/K$ at $T = 0.3t$, or for $n=1.2$ (green), $S \approx 57 \mu V/K$ at $T = 0.24t$, or for $n=1.08$ (dashed green), $S \approx 71.23 \mu V/K$ at $T = 0.01t$. This represents at least an increase by a factor 3-4 to more than an order of magnitude with respect to the non-interacting case. The appearance of pronounced peaks at low temperature that are absent in the non-interacting limit reflects, as pointed out in Ref.~[\onlinecite{Merino:2000}], the fact  that Fermi liquid coherence is recovered only at very low temperature in the presence of strong correlations.~\cite{dmft} The peaks for $S$ are similar to those of the specific heat (see Fig.9 of Ref.~[\onlinecite{Arsenault:Semon:Tremblay:2012}]) and define $T^{coh}$, as noted before~\cite{Merino:2000}. Since $T^{coh}$ in IPT-$D$ and CTQMC is similar~\cite{Arsenault:Semon:Tremblay:2012} the positions of the peaks in $S_{Kubo}$ should be correct. Frustration is a factor that contributes to make the Fermi liquid coherence $T$ low. It also plays a role in the $t-J$ model on the triangular lattice.~\cite{Haerter:Peterson:Shastry}. The results also confirm the predictions of Shastry using $S^{**}$ concerning the dependence of the thermopower on the sign of $t$ on a frustrated lattice.

The results very near half-filling, shown by dashed lines in Fig.~\ref{fig:S_Kubo_IPT}(d) are especially interesting: a) Large thermopower develops at low temperature above half-filling where there is frustration. b) The low temperature slope changes sign near $n=1$ by contrast with the smaller $U$ results where the sign change occurs near $n=0.8$. We show in Appendix~\ref{n=1} that the change of sign of $S_{Kubo}$ at low temperature near $n=1$ can be confirmed directly from the CTQMC results. From a Fermi liquid point of view this phenomenon reflects the importance of particle-hole asymmetric terms $\omega T^2$ and $\omega^3$ in the scattering rate.~\cite{Haule:Kotliar:2009}

\section{Other measures of thermopower:} Here we compare $S^{**}(T)$ and $S_{Kelvin}(T)$ respectively in Figs.~\ref{fig:S_Kubo_IPT}(b),(e) and (c),(f) with the corresponding $S_{Kubo}(T)$ . The results for $S^{**}(T)$ and $S_{Kelvin}(T)$, obtained directly with CTQMC without analytical continuation, provide both physical understanding and an alternate more straightforward way to estimate the thermopower. For $S^{**}(T)$, results for both $U = 8t$ and $U = 32t$ obtained using IPT give basically the same results as those obtained from CTQMC. Hence they can be considered reliable. For $S_{Kelvin}(T)$ at large $U = 32t$, we obtained the same qualitative behavior of the thermopower but with smaller coherence temperatures for CTQMC, as shown in Appendix~\ref{S_Kelvin32}. Above the coherence temperature, even the quantitative differences disappear.

\begin{figure}[tbp]
\begin{center}
\includegraphics[scale=0.5,bb=96   270   515   550]{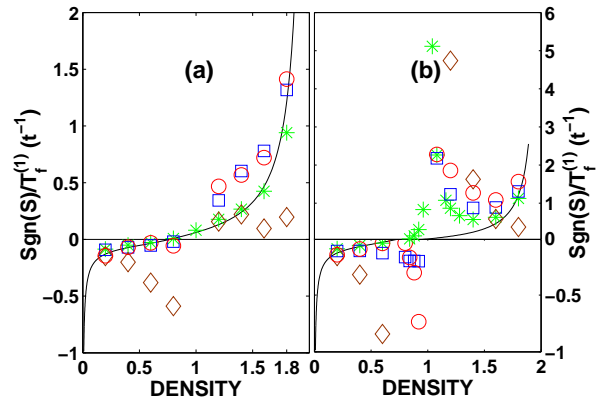}
\end{center}
\caption{(Color online) $Sgn(S)/T_f^{(1)}$ for $U=8t$ (a) and $U=32t$ (b) as a function of the density. The Boltzmann non-interacting value is given by the black solid line, $S_{Kubo}$ (blue ($\square $)), $S_{Fermi}$ with only the first term (green ($\ast$)), $S_{Kelvin}$ (brown ($\lozenge $)) and $S^{**}$ (red ($\circ $)).}
\label{fig:Tf_1}
\end{figure}

We focus first on the low $T$ linear regime, which corresponds most of the time to operating conditions. In the spirit of the Boltzmann calculation for $U=0$ in the constant relaxation-time approximation, we have a phenomenological form for $S$ at low $T$ involving two different Fermi temperature $T_f^{(1)}$ and $T_f^{(2)}$
\begin{equation}\label{S_approx}
    S \approx Sgn(S) \d{k_B}{|q_e|}\d{\pi^2}{2}\d{T}{T_f^{(1)}}\left( 1 - \left(\d{T}{T_f^{(2)}}\right)^2 \right).
\end{equation}
Depending on the band structure and filling, $T_f^{(1)}$ and $T_f^{(2)}$ can be very different from each other even at $U=0$. We wrote explicitly the sign of the thermopower to be able to define a positive $T_f^{(1)}$. Fig.~\ref{fig:Tf_1} shows $Sgn(S)/T_f^{(1)}$ as calculated from the Boltzmann non interacting $S$ (solid black line), and for $S_{Kubo}$ (blue squares). The latter is our reference value for finite $U$. For comparisons, we plot $S^{**}$ (red circles), and $S_{Kelvin}$ (brown diamonds) as well as $S_{FL}$ (green star), the value obtained by assuming that we have a Fermi liquid with a pure $\omega^2$ dependence without deviations from particle-hole symmetry, i.e. the first term of $S_{Fermi}$ in Eq.\eqref{SFermi}.~\cite{Haule:Kotliar:2009} The left panel is for $U = 8t$ (a) and the right panel for $U = 32t$ (b). This illustrates two of the important results of this paper i.e. (1) for weak to intermediate coupling in Fig.~\ref{fig:Tf_1}~(a), $S^{**}$  is a good indicator of the behavior of $S$. (2) The failing of $S_{Kelvin}$ in that region suggests that the common view of the thermopower as the entropy per particle is not to be taken seriously when the system is coherent. Still, $S_{Kelvin}$ for large $U$, Fig.~\ref{fig:Tf_1}(b), predicts the correct sign for the low $T$ thermopower. Fig.~\ref{fig:Tf_1}(b) also shows that $S^{**}(T)$ gives information about the increase of $S$ close to $n=1$ even if its wrong just below half-filling where the quasi-particles weights become very small while the scattering rates become large \cite{Arsenault:Semon:Tremblay:2012} and particle-hole asymmetric. By comparing $S_{FL}$ (green star) first term with $S_{Kubo}$ (blue squares), we finally see that at strong coupling the effects of the non-quadratic terms of the self-energy on the low $T$ thermopower are important mostly for $n < 1$.

The fact that $S^{\ast \ast }$ is less reliable for the Hubbard model at strong coupling, where interaction matters, contrary to the $t-J$ model, \cite{Haerter:Peterson:Shastry}, was expected from the fact that large $U$ makes $\omega _{c}$, the largest characteristic frequency in the problem, too far from the $\omega = 0$ limit.

Overall then, one finds from the results in Fig.~\ref{fig:S_Kubo_IPT} that for weak to intermediate coupling ($U = 8t$), the frequency dependence of the scattering rate is not so important and $S^{**}$ gives a good estimate of the thermopower. On the other hand, for strong coupling ($U = 32t$) the situation is more complicated. At low $T$, where single-site DMFT gives a Fermi liquid regime, $S^{**}$ seems to give a good estimate when the carrier density is not too close to half-filling where interactions effects become important. By contrast, when Fermi liquid behavior is present $S_{Kelvin}$ is not reliable at low temperature and strong coupling. At high $T$ however, $S_{Kelvin}$ gives qualitatively correct results,~\cite{Georges:2012} accounting for the fact that the temperatures where the thermopower changes sign are off. This demonstrates, through Eq.~\eqref{skelvin}, that in the strong coupling incoherent regime, entropy drives the thermopower. The relationship between entropy and thermopower has been discussed in the linear low temperature regime and multiband systems experimentally \cite{Skurai:1993,Behnia:2004,Sakurai:2005} and theoretically \cite{Zlatic:2007}.

Let us come back to the high temperature regime, $U\gg T \gg t$. In this limit, the so-called Mott-Heikes formula also suggests a relationship between the entropy per particle and thermopower. Heike's formula, $S \rightarrow {\mu}/{T}=(\pd s/\pd n)_{E,V}$ is derived by applying the atomic limit to the case of large $U$. The thermopower in this case depends only on density since ${\mu}/{T}$ is $T$ independent. It is given by~\cite{Chaikin:Beni:1976}
\begin{equation}\label{Beni}
  S_{MH} = \left\{
\begin{array}{rl}
-\d{k_B}{|e|}\ln\left(\d{2(1-n)}{n}\right) & \text{if } 0 \leq n \leq 1,\\
\d{k_B}{|e|}\ln\left(\d{2(n-1)}{2-n}\right) & \text{if } 1 \leq n \leq 2.
\end{array} \right.
\end{equation}
One striking prediction of these formula is a change of sign of the thermopower at $1/3$ doping from half-filling. None of our results in Fig.~\ref{fig:S_Kubo_IPT} are in the regime where ${\mu}/{T}$ is $T$ independent~\cite{Arsenault:Semon:Tremblay:2012}. It is quite remarkable then that for temperatures between $T=t$ and $T=2t$ we nevertheless have more than one sign change as a function of density for a fixed $T$.

\begin{figure}[tbp]
\begin{center}
\includegraphics[scale=0.6]{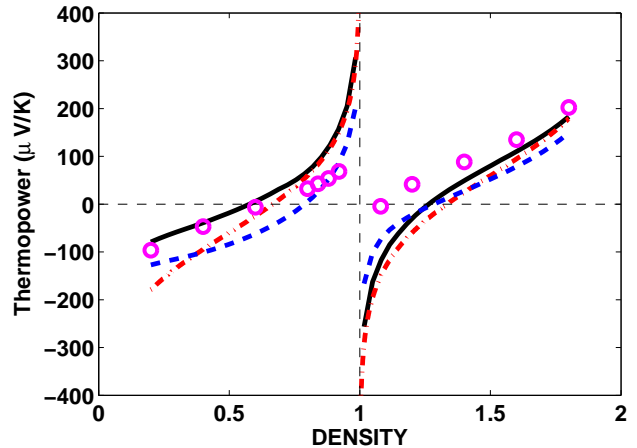}
\end{center}
\caption{(Color online) $S_{Kubo}$ (black -), $S_{Kelvin}$ (blue - -), $S_{MH}$ (red -.) and $S^{**}$ (magenta $\circ$ CTQMC) in $\protect\mu V/K$ for $U=32t$ as a function of density for $T = 2t$. In order to calculate $S_{Kelvin}$ we used IPT results. As shown in Fig.~\ref{fig:S_Kelvin_compare}, in this temperature range, IPT-$D$ gives essentially the same $\mu (T)$ as CTQMC.}
\label{graph:ST2}
\end{figure}

Even though  $S_{Kelvin}= S_{MH}$ in the regime where $\mu/T$ is $T$ independent, when $T$ is not high enough for this condition to be realized, as in our case, $S_{Kelvin}$ makes a stronger case than Heikes formula. Indeed, $S_{Kelvin}=(\pd s/\pd n)_{T,V}$ relies on the correctness of interchanging the order of two limits, which can become justified in the incoherent regime~\cite{KotliarPrivate:2012}. It was checked explicitly in Ref.~[\onlinecite{Georges:2012}] that Kelvin's formula is better than a modified Heike's formula, $(\mu (T) - \mu (0))/T$. To illustrate what happens in our case, we show in Fig.~\ref{graph:ST2} the thermopower as a function of density for $T = 2t$ and $U = 32t$ for $S_{Kubo}$,  $S_{Kelvin}$, $S^{**}$ and compare with $S_{MH}$ in Eq.~\eqref{Beni}. For densities above about $n=0.8$, our Kubo results match quite well with Heike's. For large doping, $S^{**}$ works better than $S_{Kelvin}$ or $S_{MH}$, which is another indication that large interaction effects are necessary for $S_{Kelvin}$ or $S_{MH}$ to be good approximations. Despite their simplicity, we see that the latter two formula capture the essence of the thermopower in the high temperature regime. Note that the quantitative results are a bit off, especially for $n<0.8$ where the absence of electronic frustration seems to leave more room for coherence. At this particular temperature, Heike's formula seems to be somewhat better overall than Kelvin's. However, in the presence of electronic frustration at $n>1$,  Heike's formula shows a small systematic deviation from Kubo's that does not appear for $S_{Kelvin}$.

Chakraborty \emph{et al}.\cite{Chakraborty:Galanakis:Phillips:2010} studied the case of a square lattice with nearest-neighbor hopping at large $U=20t$ for $T=0.5t$ in the incoherent regime. $U=40t$ and $T=t$ would be more comparable to our case given that the bandwidth of the fcc lattice is twice as large as that of the square lattice for the same $t$. They solved the problem using CDMFT with NCA as an impurity solver. They did the calculation for densities between $n = 0.6$ and $n=1$ and found, like we do, that Kubo's result is quite close to Heike's. A more thorough investigation of the high-temperature regime is called for.

\section{Summary}
We have shown that from weak to intermediate coupling, the thermopower of the FCC lattice is not influenced very much by interactions. The quantity $S^{\ast \ast }$, that neglects the frequency dependence of the scattering rate, can be a reliable estimator for the thermopower. By contrast, at strong coupling, namely when at $n=1$ the system is a Mott insulator, sharp peaks in the $T$ dependence of Seebeck coefficient appear at low temperature and, for the range of dopings considered, absolute values comparable to $k_{B}/|q_e|,$ close to hundreds of $\mu V/K,$ can be obtained. We found a sharp increase of the thermopower at a given $T$ when one approaches $n=1$. The increase is particularly sharp for $n>1$ where there is strong electronic frustration. We leave open the question of whether there is an optimal doping or just a sudden jump from very large thermopower at infinitesimal doping to zero at half-filling. For doped Mott insulators in the incoherent large-temperature limit, the Kelvin formula gives a reasonable estimate of the thermopower without the need for analytical continuation. This clearly shows that large thermopower in this case can be explained on entropic grounds over the whole doping range at large $T$.

It would be of great interest to verify how $S_{Kelvin}$ behaves at low $T$ and large $U$ when non-Fermi liquid behavior is observable. For this, one needs to consider lower-dimensional systems that require going beyond single-site DMFT. Also, addition of disorder trough CPA~\cite{CPA} or statistical DMFT~\cite{stat:DMFT} would be of interest to verify how $S^{**}$ and $S_{Kelvin}$ inform us about the interplay between correlations and disorder.

\begin{acknowledgments}
We acknowledge S.R. Hassan for help and discussion at the beginning stage of this work, J. Freericks, G. Kotliar, P. Phillips and W. Xu for discussions. We are grateful to Antoine Georges for sharing his insights with us before publication. A.-M.S.T and B.S.S. thank the Aspen Center for Physics where this collaboration began. B.S.S. was supported at UCSC  by DOE under Grant No. FG02-06ER46319.
We  acknowledge support from the Natural Sciences and Engineering Research Council
of Canada (NSERC) (L.-F.A. and A.-M.S.T),  and by the Tier I Canada Research Chair
Program (A.-M.S.T.). Computational resources were provided by CFI, MELS, the
RQCHP and Compute Canada.
\end{acknowledgments}
\appendix
\section{Sign change of the thermopower around $n=1$ for $U=32t$}\label{n=1}
Since the calculation of $S_{Kubo}$ rests on IPT-$D$, we need to verify that the change of sign of the low temperature slope around $n=1$ is not an artefact of the method. We present three arguments. First, we show that it is possible to obtain the change of sign purely from CTQMC results. Although analytical continuation for CTQMC is nearly impossible for such a large $U$, there is a way to check the sign of the thermopower directly from the behavior in Matsubara frequency. Following Ref.~[\onlinecite{Gunnarsson:2010}] we define the function
\begin{equation}\label{gamma}
    \gamma_{ij} (\Omega_n) \equiv \d{\chi_{ij}(0)-\chi_{ij}(i\Omega_n)}{\Omega_n} = \int_{-\infty}^{\infty}\d{d\omega}{\pi}\d{\Omega_n}{\omega^2+\Omega_n^2}L_{ij} (\omega).
\end{equation}
In Eq.~\eqref{gamma}, $i$ and $j$ can take the value 1 or 2. $\chi_{11}$ is the charge-charge, $\chi_{12}$ is the charge-heat and $\chi_{22}$ is the heat-heat susceptibility. The sign of the thermopower is thus given by the sign of $L_{12}(0)$. In terms of the $\gamma_{ij}$ function, we can show that
\begin{equation}\label{conduc_2}
    L_{ij}(0) = \lim_{\Omega_n\rightarrow 0}\gamma_{ij} (\Omega_n).
\end{equation}
If we replace the discrete variable $\Omega_n$ by a continuous variable $z$, Eq.~\eqref{conduc_2} would correspond to
\begin{equation}\label{conduc_3}
    \lim_{\Omega_n\rightarrow 0}\gamma_{ij} (\Omega_n) \rightarrow \lim_{z\rightarrow 0}\gamma_{ij} (z) = -\chi'_{ij}(z)\Big|_{z\rightarrow 0}.
\end{equation}
Thus the DC value would be given by the slope at the origin. The problem is to find the correct continuous function since the value of the derivative will depend strongly on the function. But for the sign there is no such problem. Direct inspection of the $\Omega_n$ dependence indicates what is the sign of the derivative and thus the sign of the thermopower. To verify again this result, we also calculated the susceptibility $\chi_{12}$  by the same method we used for $\chi_{11}$ in Ref.~[\onlinecite{Arsenault:Semon:Tremblay:2012}] but using results obtained via CTQMC. We find that, indeed, at $U=32t$, the low temperature thermopower changes sign around half-filling rather than around the small $U$ result $n=0.80$. Finally, we can also use just a few Matsubara frequencies to obtain the self-energy near zero frequency from Padé approximants and use it to calculate $L_{12}(0)$. The sign should be correct even though the actual value cannot be trusted. Once again we obtain that the sign change occurs around half-filling if we use 5, 10 or 20 Matsubara frequencies. We are thus confident that the IPT-$D$ results for $S_{Kubo}$ are at least qualitatively significant.

The sign change and its relation to the underlying Fermi liquid predicted by DMFT can also be understood analytically. Starting from an $\omega^2$ self-energy, Refs.~[\onlinecite{Palsson:Kotliar:1998,Haule:Kotliar:2009}], demonstrated that the thermopower only depends on interaction through $Z$, the quasiparticle weight. Hence, at first sight the sign seems determined solely by the non-interacting problem, namely by the sign of the derivative of the transport function Eq.~\ref{TransportFunction}. Howewer, when one considers the particle-hole asymmetry of the quasiparticle lifetime by including $\omega T^2$ and $\omega^3$ terms in the self-energy~\cite{Haule:Kotliar:2009}, the thermopower is given by
\begin{equation}
S_{Fermi} = -\d{k_B}{|e|}\d{k_BT}{Z}\left[ \d{\phi'(\tilde{\mu})}{\phi(\tilde{\mu})}\d{E_2^1}{E_0^1} - \d{a_1E_4^2 + a_2E_2^2}{\gamma_0E_0^1}  \right],
\label{SFermi}
\end{equation}
where $\tilde{\mu} = \mu - \text{Re}\{\Sigma (0)\}$, the $E's$ are universal constants while $a_1$, $a_2$ and $\gamma_0$ are fitting parameters for the different powers of the self-energy and thus have an interaction dependence. Recent strong coupling theories~\cite{Shastry:2012} provide an analytical understanding of the odd frequency contributions to the quasiparticle lifetimes, which become prominent as the particle density increases towards half filling. These provide dynamical asymmetry and significantly influence the ARPES line shapes. From Eq.~\eqref{SFermi}, we see that the asymmetry in the lifetime creates a term that might compete for the sign and magnitude of the thermopower. This is what is happening in our case. Below half filling but above $n=0.8$, the band results would be positive but the interacting part leads to a strong particle-hole asymmetry that competes and keeps $S_{Kubo}$ negative. This compensation leads to a small thermopower between $n=0.8$ and $n=1$. However, above half-filling, either both contributions are of the same sign and enable an increase in the thermopower or the second contribution is negligible. The asymmetry in the scattering rate also plays an important role in the temperature dependence of the thermopower, as shown in Ref.[\onlinecite{Georges:2012}].

\section{$S_{Kelvin}$ at $U=32t$}\label{S_Kelvin32}
\begin{figure}[tbph]
\begin{center}
\mbox{\includegraphics[scale=0.5]{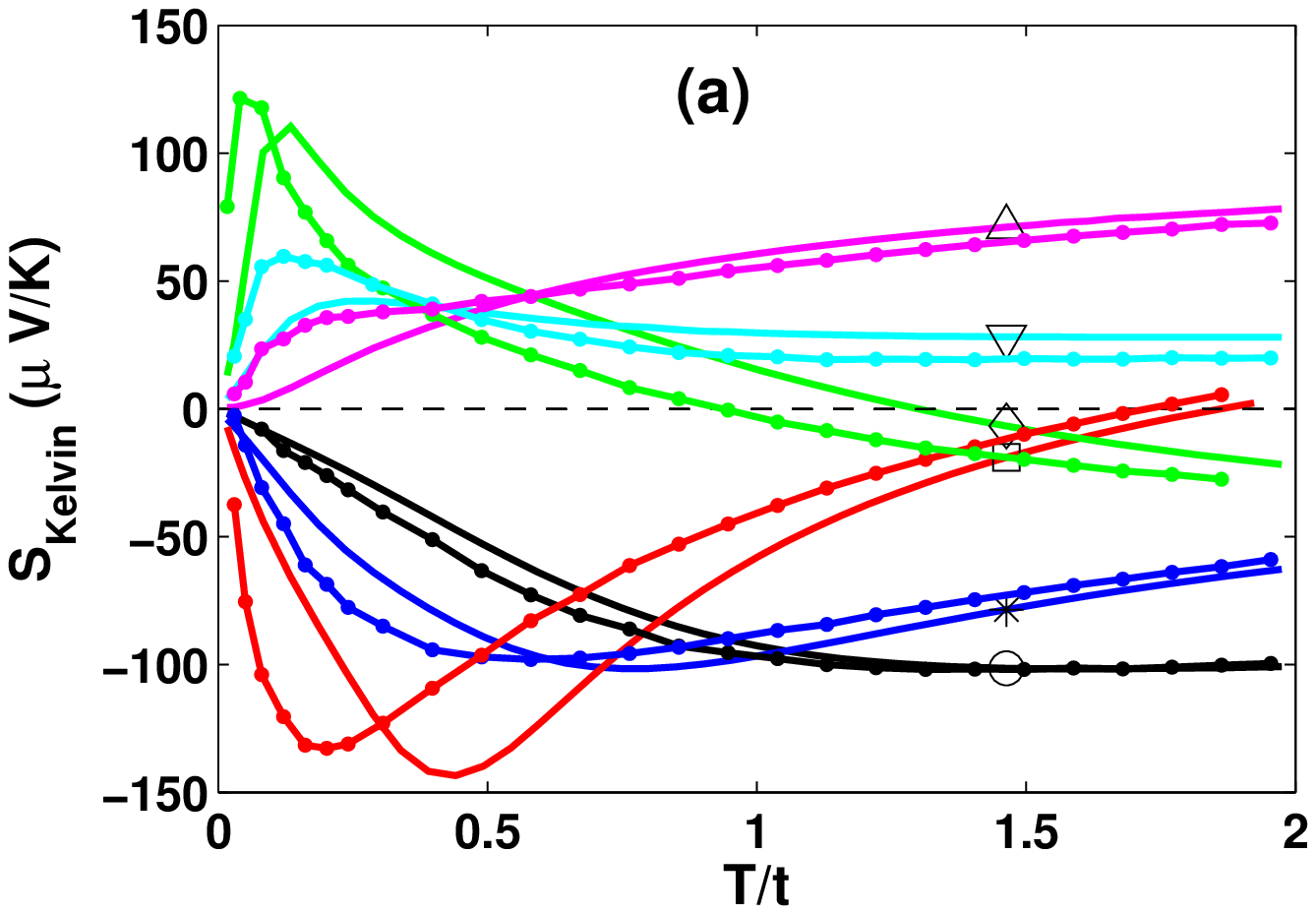}}
\mbox{\includegraphics[scale=0.5]{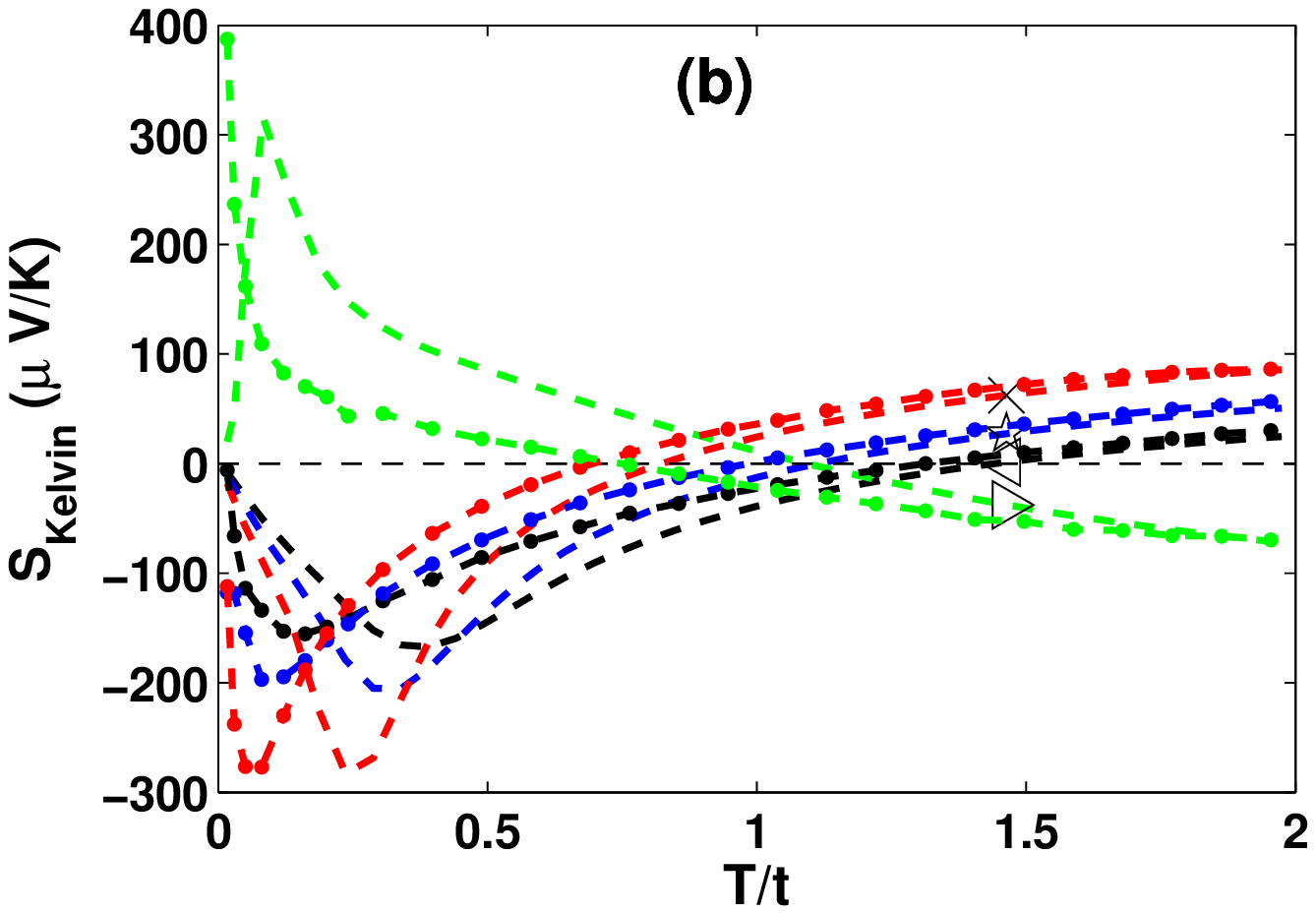}}
\end{center}
\caption{(Color online) $S_{Kelvin}$ as a function of $T/t$ as calculated with
IPT (-) and CTQMC (.-) for $U=32t$. In (a) values of filling farther away from $n=1$ and (b) fillings closest to $n=1$: $n=0.4$ (black ($\circ $)), $0.6$ (blue ($\ast$), $0.8$ (red ($\square $)), $0.84$ (black - - ($\vartriangleleft $)), $0.88$ (blue - - ($\star$)), $0.92$ (red - - ($\times $))), $1.08$ (green - - ($\triangleright$)) $1.2$ (green ($\lozenge $)), $1.4$
(cyan ($\triangledown $)) and $1.6$ (magenta ($\vartriangle $)).}
\label{fig:S_Kelvin_compare}
\end{figure}

The only quantity that does not require analytical continuation and depends on the method of calculation, i.e. IPT-$D$ vs CTQMC, is $S_{Kelvin}$ at strong coupling. The comparisons are made in Fig.~\ref{fig:S_Kelvin_compare}. Lines with the same color are obtained for he same filling. There are dots when they are obtained from CTQMC. The main quantitative difference is that the coherence temperatures (peaks) are smaller with CTQMC and closer to the ones predicted by $S_{Kubo}$ and $C_n$ even if those two quantities were obtained with IPT-$D$. This quantitative difference does not prevent qualitative comparisons. For $T$ larger than the peaks, then even the quantitative differences disappear.

\end{document}